\documentclass[12pt]{article}
\usepackage{amsfonts,amsmath}
\usepackage[hypertex]{hyperref}
\newcommand{\gt}{\tau}
\makeatletter \@addtoreset{equation}{section} \makeatother
\newcommand{\bp}{\bar{\partial}}
\def\theequation{\thesection.\arabic{equation}}

\newcommand{\be}{\begin{equation}}
\newcommand{\ee}{\end{equation}}
\newcommand{\bee}{\begin{eqnarray}}
\newcommand{\beee}{\begin{array}}
\newcommand{\eee}{\end{eqnarray}}
\newcommand{\eeee}{\end{array}}

\newcommand{\bV}{\overline{V}}

\newcommand{\ga}{\alpha}
\newcommand{\pa}{{\dot{\ga}}}
\newcommand{\pb}{{\dot{\gb}}}

\newcommand{\gb}{\beta}
\newcommand{\gga}{\gamma}

\newcommand{\G}{{\mathcal G}}

\newcommand{\Ll}{{\cal L}}
\newcommand{\Hh}{{\cal H}}

\newcommand{\rhs}{{\it r.h.s.} }

\newcommand{\ie}{{\it i.e.,} }
\newcommand{\ls}{\!\!\!\!\!\!}

\newcommand{\gd}{\delta}

\newcommand{\gvep}{\varepsilon}

\newcommand{\go}{\omega}

\newcommand{\q}{\,,\qquad}

\newcommand{\nn}{\nonumber}

\newcommand{\half}{\frac{1}{2}}

\newcommand{\p}{\partial}
\newcommand{\D}{{\cal D}}

\newcommand{\f}{\frac}

\newcommand{\bu}{\bar{\kappa}}

\def\G{\Gamma}

\def\s{\sigma}

\newcommand{\PPP}{ {J} }

\newcommand{\dgb}{{\dot \gb}}
\newcommand{\dga}{{\dot \ga}}

\newcommand{\dr}{{\rm d}}

\begin{document}

\begin{flushright}

{\small FIAN/TD/16-2017}
\end{flushright}
\vspace{1.7 cm}

\begin{center}
{\large\bf On the Local Frame in
 Nonlinear Higher-Spin Equations
}

\vspace{1 cm}

{\bf  M.A.~Vasiliev}\\
\vspace{0.5 cm}
{\it
 I.E. Tamm Department of Theoretical Physics, Lebedev Physical Institute,\\
Leninsky prospect 53, 119991, Moscow, Russia}

\end{center}

\vspace{0.4 cm}

\begin{abstract}
\noindent
Properties of the resolution operator $\dr_{loc}^*$ in higher-spin equations, that leads to local
current interactions at the cubic order and minimally nonlocal higher-order
corrections, are formulated in terms of the condition on the class of master fields of
higher-spin theory that restricts both the dependence on the spinor $Y$, $Z$ variables
and on the contractions of indices between the constituent fields in bilinear
terms. The Green function in the sector of zero-forms is found for the case of
constituent fields carrying helicities of opposite  signs. It is shown that the local
resolution  $\dr_{loc}^*$ differs from the conventional De Rham resolution  $\dr_Z^*$
by a non-local shift.

\end{abstract}

\newpage
\tableofcontents

\newpage

\section{Introduction}
Nonlinear field equations for massless fields of all spins in four dimensions were
found in \cite{Vasiliev:1990en,Vasiliev:1992av}. The most symmetric vacuum solution to
 these equations describes $AdS_4$. Due to the presence  of $AdS_4$  radius as a
  dimensionful parameter, higher-spin (HS) interactions can contain
infinite tails of higher-derivative terms. This can make the theory not local in the standard
sense, raising the question which field variables lead to the local or minimally nonlocal
setup in the perturbative analysis.

Geometric origin of the dimensionful parameter $\rho$
has an important consequence that any HS gauge theory  with
unbroken HS symmetries does not allow a low-energy analysis because
a dimensionless derivative $\rho \f{\p}{\p x}$ where $\rho=\lambda^{-1}$
is the $AdS$ radius, that appears in the
expansion in powers of derivatives, cannot be treated as small.
This is because  the rescaled covariant derivatives
 $\D=\rho D$, which are non-commutative in the background $AdS$ space-time of curvature $\rho^{-2}$,
have commutator of order one, $[\D\,,\D] \sim 1$.
As a result, all terms with higher derivatives may give comparable contributions.

Another important feature of HS theory is that it describes interactions of infinite
towers of HS fields while HS symmetry transforms a spin-$s$ field   to fields of other
spins. In particular, HS symmetries transform spin two to higher spins.
As emphasized in \cite{Vasiliev:2014vwa}, this has a consequence that such concepts of
Riemannian geometry as space-time point and dimension
are not necessarily applicable to interacting HS theories.

The importance of the proper definition of locality was originally emphasized in \cite{prok}
where it was shown that by a seemingly local field redefinition  it is possible
to get rid of the currents from the nonlinear HS field equations in $AdS_3$.
The question what is a  proper definition of a weakly local field redefinitions
in HS nonlinear theories was addressed recently in several papers
\cite{Vasiliev:2015wma,Boulanger:2015ova,Bekaert:2015tva,Skvortsov:2015lja,
Vasiliev:2016xui,Taronna:2016xrm}, in particular, in the context
 of the derivation of current interactions of massless fields from
the nonlinear HS field equations  of \cite{Vasiliev:1992av}.
In \cite{Vasiliev:2015wma}
a proposal was put forward on the part of the problem associated with the
exponential factors resulting from so-called inner Klein operators while the
structure of the preexponential factors was only partially determined. In this
paper we make a step towards  completing the definition  of
locality in the setup of HS equations of  \cite{Vasiliev:1992av}.

Conclusions of the papers \cite{Boulanger:2015ova,Skvortsov:2015lja,Taronna:2016xrm} and
\cite{Vasiliev:2016xui} were somewhat opposite.
The authors of \cite{Boulanger:2015ova,Skvortsov:2015lja,Taronna:2016xrm}
argued
that no preferred frame of field variables in HS theory exists, bringing
current interactions to the proper form with the conclusion
that using field redefinitions
exhibiting the same asymptotic behaviour  it is possible to obtain current interactions with
arbitrary charges in front of different currents, including zero charges which means no interactions. In the absence of a further selection criterion
 such a conclusion would imply  difficulties of the physical interpretation of the HS equations of \cite{Vasiliev:1992av}.

The analysis of \cite{Boulanger:2015ova,Skvortsov:2015lja,Taronna:2016xrm}
was performed in the one-form gauge sector of equations of \cite{Vasiliev:1992av}.
On the other hand, in \cite{Vasiliev:2016xui} we considered the problem in the zero-form
sector of the HS equations which is simpler in many respects being the same time fully
informative, finding a simple field redefinition that brings the quadratic corrections
to free field equations following from the nonlinear HS  equations to the canonical
local current form in agreement with the unfolded form of current interactions found
earlier in \cite{Gelfond:2010pm}. In \cite{Gelfond:2017wrh} these results were extended to the
one-form sector. The results of \cite{Vasiliev:2016xui} were then shown in
\cite{Didenko:2017lsn} (for a related particular result  see also \cite{Sezgin:2017jgm}) to lead to correct $AdS/CFT$
predictions at the boundary of $AdS_4$ thus resolving some of the puzzles of the
analysis of HS holography conjectures of \cite{Klebanov:2002ja,Aharony:2011jz}
encountered in  \cite{Giombi:2009wh,Giombi:2011kc,Giombi:2012ms}
(and references therein).

The resulting couplings for different HS currents were expressed unambiguously in
\cite{Vasiliev:2016xui,Gelfond:2017wrh}
via the coupling constant of the nonlinear HS equations. In this paper it will be shown that
the choice of field variables of \cite{Vasiliev:2016xui} is distinguished by the
condition that the associated higher-order corrections in HS theory are minimally nonlocal.
To see this it is necessary to take into account the dependence on the spinor
variables  $Z^A$ in addition to that on $Y^A$
considered in \cite{Boulanger:2015ova,Skvortsov:2015lja,Taronna:2016xrm},
that is insufficient to  control higher-order locality.
It will be shown that
upon a proper redefinition of the resolution   underlying the procedure of
solving the nonlinear HS equations from the conventional one $\dr^*_Z$ to $\dr^*_{loc}$
the results of \cite{Vasiliev:2016xui} come out directly with no need of further field
redefinitions.  This implies that the  De Rham resolution
$\dr^*_Z$ is related to the local one $\dr^*_{loc}$ by a
 nonlocal field redefinition.

In this paper we determine explicit form of $\dr^*_{loc}$ in the sector of zero-forms
and to the lowest order. Extension of these results to  forms of higher degrees
 and to higher orders will be given elsewhere \cite{hom}.

It should be stressed that, as explained in more detail below, the freedom in the
definition of the resolution
parametrizes the freedom in the choice of the homogeneous part of the solution to the
differential equations in spinorial $Z$ variables in the process of solving nonlinear
HS equations. As usual, this freedom has to be fixed by appropriate boundary conditions.
In quantum mechanics proper boundary conditions for solutions to the Schrodinger equation
are fixed from the  normalizability condition.
(Discarding the normalizability condition may lead to meaningless conclusions, making
it impossible to find a physical spectrum in a problem in question.)
In HS theory, the appropriate choice of the boundary conditions is determined
by the minimal nonlocality condition (which in the cases considered so far
implies locality of HS vertices). Since, as explained below, this
 demands the solution to belong to a proper functional class, it is
natural to speculate that, eventually, this functional class can be determined
by a certain normalizability condition. The results of this paper provide a
starting point for  exploring  this problem setting.

 The rest of the paper is organized as follows. In Section \ref{CMT}
 we recall unfolded formulation of free massless fields in four dimensions.
 The form of nonlinear HS equations is
 sketched in Section \ref{Nonlinear Higher-Spin Equations}.
 In Section \ref{peran} details of the perturbative analysis of nonlinear
 equations with some emphasize on the homotopy technique are presented.
 The form of the local resolution  $\dr^*_{loc}$  leading to proper current
 interactions is found in Section \ref{loc}. The structure of the unfolded equations
 describing current interactions of massless fields of all spins
 in the zero-form sector is recalled in Section \ref{y}. Green function for
 the zero-form sector of HS equations is found in Section \ref{gf0} for the case of
  opposite helicity signs of the
 constituent fields of the current. Conclusions and perspective are briefly discussed
 in Section \ref{disc}. Appendix contains details of the derivation of the Green function.

\section{Free fields}
\label{CMT}

In the frame-like formulation, the infinite set of $4d$ Fronsdal \cite{Frhs,Frfhs}  massless fields of all spins $s=0,1,2\ldots $ is described  by a
{one-form} $  \omega (Y;K| x)= dx^n\omega_n (Y;K| x)$ \cite{Vasiliev:1980as,Vasiliev:1986td} and zero-form
$ C(Y;K| x)  $ \cite{Ann}
\be
\label{f}
 f(Y;K| x) =\f{1}{2i}\sum_{n,m=0}^{\infty}\sum_{i,j=1}^2 \frac{1}{n!m!}
{y}_{\alpha_1}\ldots {y}_{\alpha_n}{\bar{y}}_{{\pb}_1}\ldots
{\bar{y}}_{{\pb}_m } k^i \bar k^j f_{ij}{}^{\alpha_1\ldots\alpha_n}{}_,{}^{{\pb}_1
\ldots{\pb}_m}(x)\,,
\ee
where $x^n$ are $4d$ space-time coordinates,  $Y^A=(y^\ga, \bar y^\dga)$
are auxiliary commuting spinor variables ($A=1,\ldots 4$ is a Majorana spinorial index
while $\ga = 1,2$ and $\dga =1,2$ are two-component ones) and
the  Klein operators $K=(k,\bar k)$ satisfy
\be
\label{kk}
k y^\ga = -y^\ga k\,,\quad
k \bar y^\pa = \bar y^\pa k\,,\quad
\bar k y^\ga = y^\ga \bar k\,,\quad
\bar k \bar y^\pa = -\bar y^\pa \bar k\,,\quad k^2=\bar k^2 = 1\,,\quad
k\bar k = \bar k k\,.
\ee
More precisely, to describe massless fields, the one-form $  \omega (Y;K| x)$
should be even in $k,\bar k$ (\ie $i=j$ in (\ref{f})) while the zero-form $C (Y;K| x)$
should be odd (\ie $i+j=1$).

The  Central on-shell theorem states that unfolded system of field equations
for free massless
fields of all spins has the form \cite{Ann}
\bee
\label{CON1}
    && \ls\ls\ls R_1(Y;K| x) = L(w,C):=
\f{i}{4} \Big ( \eta \overline{H}^{\dga\pb}\f{\p^2}{\p \overline{y}^{\dga} \p \overline{y}^{\dgb}}\
{ C}(0,\overline{y};K| x) k +\bar \eta H^{\ga\gb} \f{\p^2}{\p
{y}^{\ga} \p {y}^{\gb}}\
{C}(y,0;K| x) \bar k\Big ) \,, \\\label{CON2}
\,&& \tilde{D}C (y,\overline{y}| x) =0\,, \eee
where
\be
\label{RRR}
R_1 (Y;K| x) :=D^{ad}\omega_1(Y;K |x) :=
D^L \omega_1 (Y;K| x) +
\lambda h^{\ga\pb}\Big (y_\ga \frac{\partial}{\partial \bar{y}^\pb}
+ \frac{\partial}{\partial {y}^\ga}\bar{y}_\pb\Big )
\omega_1 (Y;K | x) \,,
\ee
\be
\label{tw}
\tilde D C(Y;K |x) :=
D^L C (Y;K |x) -{i}\lambda h^{\ga\pb}
\Big (y_\ga \bar{y}_\pb -\frac{\partial^2}{\partial y^\ga
\partial \bar{y}^\pb}\Big ) C (Y;K |x)\,,
\ee
\be
\label{dlor}
D^L f (Y;K|x) :=
\dr f (Y;K|x) +
\Big (\go_0^{\ga\gb}y_\ga \frac{\partial}{\partial {y}^\gb} +
\overline{\go}_0^{\pa\pb}\bar{y}_\pa \frac{\partial}{\partial \bar{y}^\pb} \Big )f (Y;K|x)\q
\dr =dx^n\f{\p}{\p x^n}
\,.
\ee
$\go_1$ describes first-order fluctuations of the
HS gauge fields.
Background $AdS_4$ space of radius $\lambda^{-1}=\rho$
is described by a flat $sp(4)$
connection $w=(\omega_{0\alpha \gb},\overline{\omega}_{0\dga\dgb},h_{\ga\dgb})$
containing Lorentz connection
$\omega_{0\alpha \gb} $, $\overline{\omega}_{0\dga\dgb}$ and
vierbein  $h_{\ga\dgb}$ that obey the  equations
\be
\label{adsfl}
R_{\ga\gb}=0\,,\quad \overline{R}_{\pa\pb}=0\,,
\quad R_{\ga\pa}=0\,,
\ee
where, here and after discarding the wedge product symbols,
\be
\label{nR}
R_{\alpha \gb}:=\dr\omega_{0\alpha \gb} +\omega_{0\alpha}{}_\gamma
 \omega_{0\gb}{}^{\gamma} -\lambda^2\, H_{\alpha \gb}\q
\overline{R}_{{\pa} {\pb}}
:=\dr\overline{\omega}_{{0\pa}
{\pb}} +\overline{\omega}_{0{\pa}}{}_{\dot{\gamma}}
 \overline{\omega}_{0{\pb}}{}^{ \dot{\gga}} -\lambda^2\,
 \overline H_{{\pa\pb}}\,,
\ee
\begin{equation}
\label{nr}
R_{\alpha {\pb}} :=\dr h_{\alpha{\pb}} +\omega_{0\alpha}{}_\gamma
h^{\gamma}{}_{\pb} +\overline{\omega}_{0{\pb}}{}_{\dot{\delta}}
 h_{\alpha}{}^{\dot{\delta}}\,.
\end{equation}
Two-component indices are raised and lowered
by $\varepsilon_{\ga\gb}=-\varepsilon_{\gb\ga}$, $\varepsilon_{12}=1$: $A^\ga =\gvep^{\ga\gb} A_\gb$,
$A_\ga = A^\gb\gvep_{\gb\ga}$ and analogously for dotted indices.
$H^{\ga\gb}=H^{\gb\ga}$ and $\overline{H}^{\pa\pb} =
\overline{H}^{\pb\pa}$ are the basis two-forms
\be
\label{H}
H^{\ga\gb} := h^{\ga}{}^\pa  h^\gb{}_\pa\,,\qquad
\overline {H}^{\pa\pb} := h^{\ga}{}^\pa h_{\ga}{}^{\pb}\,.
\ee
$\eta$ and $\bar \eta$ are free complex conjugated  parameters.

Due to the dependence on the Klein operators the one-form HS connection $  \omega (Y;K | x)$
contains a doubled set of HS gauge fields. For spins $s\geq 1$, equation (\ref{CON1})
expresses the Weyl {0-forms} $C(Y;K|x)$ via
gauge invariant combinations of derivatives of the HS gauge connections.
More precisely, the primary-like Weyl {0-forms} are just the holomorphic and antiholomorphic
parts $C(y,0;K|x)$ and $C(0,\bar y;K|x)$ which appear on the
\rhs of Eq.~(\ref{CON1}).

$C(y,0;K| x)$ and $C(0,\bar y;K| x)$ describe gauge invariant combinations of derivatives of
the gauge fields of spins $s\geq 1 $ and the matter fields of spins $s=0$ or $1/2$.
For $s=1,2$, $C(y,0;K| x)$ and $C(0,\bar y;K| x)$ parameterize Maxwell and Weyl
tensors respectively. Those associated with higher powers of auxiliary variables
$y$ and $\bar y$ describe on-shell nontrivial combinations of derivatives
of the generalized Weyl tensors as is obvious from equations (\ref{CON2}), (\ref{tw})
relating second derivatives in $y,\bar y$ to
the $x$ derivatives of  $C (Y;K|x)$ of lower degrees
in $Y$. Higher derivatives in the nonlinear system result from the
 components of $C (Y;K| x)$ of higher degrees in $Y$.
$AdS$ geometry induces filtration
with respect to space-time derivatives rather than  gradation as would be
 the case for massless fields in Minkowski space free of a dimensional parameter.

System (\ref{CON1}), (\ref{CON2}) decomposes into  subsystems
of different spins, with a massless spin $s$  described by
the one-forms $ \omega (y,\bar{y};K| x)$ and zero-form $C (y,\bar{y};K| x)$ obeying
the homogeneity conditions
\be
\omega (\mu y,\mu \bar{y};K\mid x) = \mu^{2(s-1)} \omega (y,\bar{y};K\mid x)\q
C (\mu y,\mu^{-1}\bar{y};K\mid x) = \mu^{\pm 2 s}C (y,\bar{y};K\mid x)\,,
\ee
where  $+$ and $-$ signs  correspond to selfdual and anti-selfdual parts
of the generalized Weyl tensors $C (y,\bar{y}| x)$ with helicities $h=\pm s$.

\section{Nonlinear higher-spin equations in $AdS_4$}
\label{Nonlinear Higher-Spin Equations}

The master fields of the construction of nonlinear equations of
\cite{Vasiliev:1992av} consist of
the zero-form $B(Z;Y;K|x)$, space-time one-form $W(Z;Y;K|x)$ and an additional spinor field
$ S_A (Z;Y;K|x)$. It is convenient to introduce anticommuting $Z-$differentials $\theta^A$, $\theta^A \theta^B=-\theta^B
\theta^A$, to interpret $S_A (Z;Y;K|x)$ as a one-form in $Z$ direction,
\be
S=\theta^A S_A (Z;Y;K|x) \,.
\ee

 HS equations determining  dependence on the
 variables $Z_A$ in terms of ``initial data"
\be
\label{inda}
\go(Y;K|x)=W(0;Y;K|x)\,,\qquad C(Y;K|x)= B(0;Y;K|x)\,,
\ee
 are formulated in terms of the
associative star product $*$ acting on functions of two
spinor variables
\be
\label{star2}
(f*g)(Z;Y)=
\int \f{d^{4} U\,d^{4}V}{(2\pi)^{4}}  \exp{[iU^A V^B C_{AB}]}\, f(Z+U;Y+U)
g(Z-V;Y+V) \,,
\ee
where
$C_{AB}=(\epsilon_{\ga\gb}, \bar \epsilon_{\dga\dgb})$
is the $4d$ charge conjugation matrix and
$ U^A $, $ V^B $ are real integration variables.
 1 is a unit element of the star-product
algebra, \ie $f*1 = 1*f =f\,.$ Star product
(\ref{star2}) provides a particular
realization of the Weyl algebra
\be
[Y_A,Y_B]_*=-[Z_A,Z_B ]_*=2iC_{AB}\,,\qquad
[Y_A,Z_B]_*=0\q [a,b]_*:=a*b-b*a\,.
\ee

The  Klein operators satisfy
\bee
\label{kk}
&&k* w^\ga = -w^\ga *k\,,\quad
k * \bar w^\pa = \bar w^\pa *k\,,\quad
\bar k *w^\ga = w^\ga *\bar k\,,\quad
\bar k *\bar w^\pa = -\bar w^\pa *\bar k\,, \\  && k*k=\bar k*\bar k = 1\,,\quad
k*\bar k = \bar k* k\,
\eee
 with
 $w^\ga= (y^\ga, z^\ga, \theta^\ga )$, $\bar w^\pa =
(\bar y^\pa, \bar z^\pa, \bar \theta^\pa )$. These relations extend the action of the star product to the
Klein operators.

The nonlinear HS equations are \cite{Vasiliev:1992av}
\be
\label{dW}
\dr W+W*W=0\,,\qquad
\ee
\be
\label{dB}
\dr B+W*B-B*W=0\,,\qquad
\ee
\be
\label{dS}
\dr S+W*S+S*W=0\,,
\ee
\be
\label{SB}
S*B=B*S\,,
\ee
\be
\label{SS}
S*S= i (\theta^A \theta_A + \theta^\ga \theta_\ga  F_*(B)* k*\kappa +
\bar \theta^\dga \bar \theta_\dga \bar F_*(B)* \bar k* \bu)
\,,
\ee
where
$F_*(B) $ is some star-product function of the field $B$.
The simplest choice  of the linear function
\be
\label{etaB}
F_*(B)=\eta B \q \bar F_* (B) = \bar\eta B\,,
\ee
where $\eta$ is a complex parameter
\be
\label{theta}
\eta = |\eta |\exp{i\varphi}\q \varphi \in [0,\pi)\,,
\ee
leads to a class of pairwise nonequivalent nonlinear HS
theories. The  cases
of $\varphi=0$ and $\varphi =\f{\pi}{2}$  correspond
to so called $A$ and $B$ HS models distinguished
by the property that they  respect parity \cite{Sezgin:2003pt}.

The left and right inner Klein operators
\be
\label{kk4}
\kappa :=\exp i z_\ga y^\ga\,,\qquad
\bu :=\exp i \bar{z}_\dga \bar{y}^\dga\,,
\ee
 which enter Eq.~(\ref{SS}), change a sign of
 undotted and dotted spinors, respectively,
\be
\label{uf}
\!(\kappa *f)(z,\!\bar{z};y,\!\bar{y})\!=\!\exp{i z_\ga y^\ga }\,\!
f(y,\!\bar{z};z,\!\bar{y}) ,\quad\! (\bu
*f)(z,\!\bar{z};y,\!\bar{y})\!=\!\exp{i \bar{z}_\dga \bar{y}^\dga
}\,\! f(z,\!\bar{y};y,\!\bar{z}) ,
\ee
\be
\label{[uf]}
\kappa *f(z,\bar{z};y,\bar{y})=f(-z,\bar{z};-y,\bar{y})*\kappa\,,\quad
\bu *f(z,\bar{z};y,\bar{y})=f(z,-\bar{z};y,-\bar{y})*\bu\,,
\ee
\be
\kappa *\kappa =\bu *\bu =1\q \kappa *\bu = \bu*\kappa\,.
\ee

\section{Perturbative analysis and resolution operator}
\label{peran}

\subsection{Perturbations}
Perturbative analysis of  Eqs.~(\ref{dW})-(\ref{SS}) assumes their
linearization around some vacuum solution.
The simplest choice is
\be
W_0(Z;Y;K|x)= w(Y|x)\q S_0(Z;Y;K|x) = \theta^A Z_A\q B_0(Z;Y;K|x)=0\,,
\ee
where $w(Y|x)$ is some solution to the flatness condition
$\dr w + w*w=0$.
A flat connection $w(Y|x)$ bilinear in $Y^A$ describes $AdS_4$
\be
\label{ads}
w(Y|x) = -\f{i}{4} w^{AB}(x) Y_A Y_B = -\f{i}{4} (\go^{AB}(x) + h^{AB}(x)) Y_A Y_B\,,
\ee
\be
\go^{AB}(x) Y_A Y_B  := \go^{\ga\gb}(x) y_\ga y_\gb + \bar \go^{\dga\dgb}(x)
\bar y_\dga \bar y_\dgb \q h^{AB}(x) Y_A Y_B  := 2h^{\ga\dgb}(x) y_\ga \bar y_\dgb\,.
\ee

Decomposing  fields  with respect to the Klein operator parity,
 $A^\pm(Z;Y;K|x)=\pm A^\pm(Z;Y;-K|x)$, HS gauge fields are
 $W^+,S^+$ and $B^-$ while  $W^-$, $S^-$ and $B^+$
 describe an infinite tower of topological fields
 with every $AdS_4$ irreducible field describing at
most a finite number of degrees of freedom. (For more detail see
\cite{Vasiliev:1992av,Vasiliev:1999ba}). They can be
treated as representing an infinite set of  coupling constants in HS
theory. In this paper all of these fields are truncated away.

The perturbative analysis goes as follows. Suppose that an order-$n$ solution has been found
\be
W= W_0 +W^{(n)}\q W^{(n)}= \sum_{k=1}^n W_k\,,
\ee
\be
S= S_0 +S^{(n)}\q S^{(n)}= \sum_{k=1}^n S_k\,,
\ee
\be
B^{(n)}= \sum_{k=1}^n B_k\,.
\ee

Then Eq.~(\ref{SB}) gives
\be
\label{[sb]}
[S_0\,, B_{n+1}]_* = -\sum_{k+l=n+1} [S_{k}\,, B_l]_*\,.
\ee

Using that $S_0$ has a trivial star-commutator with the Klein operators $K$, an elementary computation
gives
\be
[S_0\,, F(Y;Z;K|x)]_* = -2i \dr_Z  F(Y;Z;K|x)\,,
\ee
where
\be
\dr_Z = \theta^A \f{\p}{\p Z^A}\,
\ee
is the $Z$-space exterior derivative.
As a result, Eq.~(\ref{[sb]})  is equivalent to
\be
\label{B+1}
\dr_Z  B_{n+1}  = -\f{i}{2}\sum_{k+l=n+1} [S_{k}\,, B_l]_*\,.
\ee
Since $B$ is a zero-form, in the first nontrivial order this gives
\be
\label{B1C}
B_1(Z;Y;K|x) = C(Y;K|x)\,,
\ee
\ie  the zero-form $C(Y;K|x)$ of the free theory appears as $Z$-space De Rham cohomology.

Analogously, Eq.~(\ref{SS}) reads as
\be
\label{S+1}
\dr_Z  S_{n+1}  = -\half \Big (i \sum_{k+l=n+1} S_k*S_l +
\eta \theta^\ga \theta_\ga  B_{n+1} * k*\kappa + \bar \eta
\bar \theta^\dga \bar \theta_\dga  B_{n+1} * \bar k* \bu \Big )\,.
\ee
In the lowest order this gives the equation
\be
\dr_Z  S_{1}  = -\half \Big (
\eta \theta^\ga \theta_\ga  C * k*\kappa + \bar \eta
\bar \theta^\dga \bar \theta_\dga  C * \bar k* \bu \Big )\,
\ee
expressing $S_1$ in terms of $C$.

In the $Z-$ space sector, perturbative analysis consists of  solving
repeatedly  the equations of the form
\be
\label{fg}
\dr_Z f(Z;Y;K|x) = J(Z;Y;K|x)\,.
\ee
 Formal consistency of  HS equations (\ref{SS}) and (\ref{SB}) guarantees that
  $J(Z;Y;K|x)$ is $\dr_Z$-closed
 \be
 \label{dJ}
 \dr_Z J(Z;Y;K|x)=0\,,
 \ee
 implying formal consistency of Eq.~(\ref{fg}). However, it admits a solution
 only if $J$ is $\dr_Z$-exact.

 Analysis of the equations involving space-time one-forms is analogous. Firstly one resolves
 all equations that contain $\dr_Z$. The remaining equations in $\dr_Z$-cohomology
 produce dynamical equations on the dynamical fields $\go$ and $C$ which reproduce
 Central on-shell theorem (\ref{RRR}), (\ref{tw}) along with all nonlinear
 corrections.

\subsection{Homotopy trick}
 Let us now recall the standard homotopy trick. Let $\dr$ be  a differential
(later on to be identified with $\dr_Z$) obeying
 \be
 \label{eq:dd_0}
 \dr^2=0
 \ee
 as well as a homotopy operator $\partial$
\begin{equation}
\partial^{2}=0.\label{eq:dwdw_0}
\end{equation}
Then the operator
\begin{equation}
A:=\{\mathrm{d}\,,\partial\}\label{eq:A_d_d}
\end{equation}
obeys
\begin{equation}
[\mathrm{d}\,,A]=0\,,\qquad[\partial\,,A]=0
\end{equation}
as a consequence of \eqref{eq:dd_0}, \eqref{eq:dwdw_0}. For diagonalizable
$A$ the standard Homotopy Lemma states that cohomology of $\mathrm{d}$,
denoted  $H(\mathrm{d})$, is  in the kernel of $A$
\begin{equation}
H(\mathrm{d})\subset KerA\,.\label{eq:hom_lemma}
\end{equation}
In this case the projector $\hat{h}$ to $KerA$
\begin{equation}
\hat{h}^{2}=\hat{h}
\end{equation}
can be defined to obey
\begin{equation}
[\hat{h}\,,\mathrm{d}]=[\hat{h}\,,\partial]=0\,.
\end{equation}

Also we can introduce the operator $A^{*}$ such that
\begin{equation}
A^{*}A=AA^{*}=Id-\hat{h}\,.\label{eq:AA*_Id}
\end{equation}
This allows us to define the {\it resolution operator}
\begin{equation}
\mathrm{d}^{*}:=A^{*}\partial=\partial A^{*}
\end{equation}
that obeys
\begin{equation}
\mathrm{d}^{*}\mathrm{d}+\mathrm{d}\mathrm{d}^{*}=Id-\hat{h}\,,
\end{equation}
which is equivalent to the resolution of identity
$
\left\{ \mathrm{d}\,,\mathrm{d}^{*}\right\} +\hat{h}=Id\,.\label{eq:res_id_gen}
$
This relation provides a general solution to the equation
\begin{equation}
\mathrm{d}f=J\label{eq:df_J-1}
\end{equation}
with $\mathrm{d}$-closed $J$ outside of $H\left(\mathrm{d}\right)$,
\ie obeying  $\hat{h}J=0$,
\begin{equation}
J=\mathrm{d}\mathrm{d}^{*}J\,.
\end{equation}
Hence
\begin{equation}
f=\mathrm{d}^{*}J+\mathrm{d}\epsilon+g,\label{eq:gen_sol}
\end{equation}
where an exact part $\mathrm{d}\epsilon$  and $g\in H\left(\mathrm{d}\right)$
remain undetermined.

The simplest choice of the homotopy operator for the exterior differential $\mathrm{d}=\mathrm{d}_{Z}$ is
\begin{equation}
\label{p}
\partial=Z^{A}\frac{\partial}{\partial\theta^{A}}\,.
\end{equation}
This gives
\begin{eqnarray}
 &  & A=\theta^{A}\frac{\partial}{\partial\theta^{A}}+Z^{A}\frac{\partial}{\partial Z^{A}},\\
\nonumber \\
 &  & A^{*}f\left(Z;Y;\theta\right)=
 \intop_{0}^{1}dt\dfrac{1}{t}f\left(tZ;Y;t\theta\right)\label{hom}.
\end{eqnarray}
Eq.~\eqref{eq:AA*_Id} is checked by using
\begin{equation}
t\dfrac{\partial}{\partial t}f\left(tx\right)=x\dfrac{\partial}{\partial x}f\left(tx\right).\label{eq:A_tx}
\end{equation}

$KerA$ consists of $Z$, $\theta$-independent functions and
thus, by Poincar\'e lemma, relation \eqref{eq:hom_lemma}
becomes an exact equality
\begin{equation}
H(\mathrm{d}_{Z})=KerA\,.
\end{equation}
Correspondingly,
\begin{equation}
\hat{h}J\left(Z;Y;\theta\right)=J\left(0;Y;0\right),
\end{equation}
while
\begin{equation}
\mathrm{d}_{Z}^{*}J\left(Z;Y;\theta\right)=Z^{A}\dfrac{\partial}{\partial\theta^{A}}\intop_{0}^{1}dt\dfrac{1}{t}J\left(tZ;Y;t\theta\right).
\label{eq:dz*}
\end{equation}

In this construction,  the   freedom
in the choice of the homotopy operator $\p$ affects the
homogeneous solution to equation (\ref{eq:df_J-1}),
\ie $\mathrm{d}\epsilon+g$. Since dynamical fields $C$ and $\go$ are valued in
the  $\dr_Z$ cohomology, this implies that going to a different $\p$ may imply
a field redefinition. Alternatively, one can directly redefine the resolution
$\dr^*$ by adding a closed form.

Though the conventional resolution operator $\dr^*_Z$ looks  simple and natural, it is
known since \cite{prok} to lead to nonlocalities at the nonlinear order.
In the next section we identify a resolution operator $\dr^*_{loc}$ that
 leads directly to the local
setup in the process of solving HS equations in the lowest nonlinear order. Its form is
deduced from the results of \cite{Vasiliev:2016xui}.
In accordance with the general analysis  the difference between the two
approaches effectively results in a field redefinition associated with the difference
between  cohomological $g$-terms in the respective
formulae (\ref{eq:gen_sol}).\footnote{Retrospectively, it should be noted that the formulation
originally found in \cite{Vasiliev:1990en} needed a field redefinition containing
higher derivatives  even at the free field level just because from the point of view
of the later formulation of \cite{Vasiliev:1992av}
it corresponded to the alternative choice of the homotopy operator $\p^\pm =
(Z\pm Y)^{A}\frac{\partial}{\partial\theta^{A}}$.}

Since $\dr^*_{loc}$ gives  a local
result while  $\dr^*_{Z}$ leads to a nonlocal one, they should  differ by  nonlocal
 cohomological $g$-terms in  (\ref{eq:gen_sol}).
One can argue following the authors of \cite{Boulanger:2015ova,Skvortsov:2015lja}
  that one can equally well choose
another resolution operator $\dr^{*\prime}_{loc}\neq \dr^*_{loc}$,
that might lead to another
 local result.  The weak point of this argument is  however due to
 insufficient representativity of the lowest-order analysis of
 consistency  insensitive to the specific coefficients in front of
different vertices. Things do change drastically in the higher orders.
As will be explained in the next section the choice of $\dr^*_{loc}$ associated with the solution found in \cite{Vasiliev:2016xui} is singled out by the
condition that higher-order corrections remain minimally nonlocal (if nonlocal at all)
which phenomenon is properly captured by the lowest-order corrections once the
dependence on $Z$ variables is taken into account.
The analysis of \cite{Vasiliev:2016xui} was based on
the separation of variables taking into account that, in accordance with the fact that the left and right
parts of $\dr_Z$ form a bicomplex, in the left(right) sector the dependence on
the right(left) spinors
remains unchanged in the lowest order being governed by the original star product.
At this condition the solution of  \cite{Vasiliev:2016xui} is
the only  one leading to local current interactions in the $Y,x$ sector.

\section{Perturbative analysis of locality}
\label{loc}

To see what distinguishes between the local frame  of HS equations of
\cite{Vasiliev:2016xui}  and other possible frames let us consider in more detail
the perturbative analysis in the sector of zero-forms  starting from De Rham resolution
$\dr^*_Z$.

The first step is to evaluate the first correction to $S$, \ie $S_1$.
Using (\ref{uf}) and (\ref{[uf]}), Eq.~(\ref{hom})
gives
\be
S_1 = S_{1\eta} + S_{1\bar\eta}\,,
\ee
where
\be
\label{S1}
S_{1\eta}(Z;Y;K|x) = - \eta z^\gb \theta_\gb \int_0^1d\tau \tau \exp (i \tau 
z_\ga y^\ga ) C(-\tau z, \bar y;K|x)*k\,,
\ee
\be
\label{bS1}
S_{1\bar\eta}(Z;Y;K|x) = - \bar \eta \bar z^\dgb \bar \theta_\dgb \int_0^1d\tau
\tau \exp (i \tau 
\bar z_\dga \bar y^\dga ) C(y, -\tau \bar z;K|x)*\bar k\,.
\ee

Resolution of (\ref{B+1}) with the homotopy operator $\p_Z$ (\ref{p})
gives for $B_2$ in the $\eta$-sector
\bee
B_{2\eta} &&\ls= -\f{i}{2} \dr^*_Z ([S_1\,, C]_*) = -\f{i}{2}\eta \int d_+^4\tau \delta (1-\sum^4_{i=1}\tau_i)
\int d^4 U d^4 V\exp i(U_A V^A +(1-\tau_3) z_\ga y^\ga)\nn\\
&&
\Big [ y^\gb u_\gb \delta(\tau_1) C(\tau_3 y -\tau_1 z +(\tau_3+\tau_4) u,\bar y+\bar u;K)
C(\tau_3 y+\tau_2 z +v,\bar y +\bar v;K)\nn\\
&&-
 y^\gb v_\gb \delta(\tau _2) C(\tau_3 y -\tau_1 z + u,\bar y+\bar u;K)
C(\tau_3 y+\tau_2 z +(\tau_3+\tau_4)v,  \bar y +\bar v) \Big ] *k*\kappa \,,
\label{bn}
\eee
where we use notation
\be
d_+^4\tau := d\tau_1 d\tau_2 d\tau_3 d\tau_4\theta(\tau_1)\theta(\tau_2)\theta(\tau_3)\theta(\tau_4)\q
\theta(\tau) = 1(0) \quad \mbox{if} \quad \tau\geq 0 (\tau<0)\,
\ee
with the convention that
\be
\theta (\tau) \delta(\tau) = \delta(\tau)\,.
\ee
Equivalently, this expression can be written in the differential form
symmetric with respect the first
and second factors of $C$
\be
\label{B2}
 B_{2\eta} = \f{1}{2}\eta \int d_+^4\tau \delta (1-\sum^4_{i=1}\tau_i)
y^\gb \big (\delta(\tau_1) \p_{2\gb}+\delta(\tau_2)  \p_{1\gb} \big) \exp (X)
 C(Y_1;K)C(Y_2;K) \Big \vert_{Y_{1,2}=0} *k*\kappa \,,
\ee
where
\be
X=  i(1-\tau_3) z_\ga y^\ga +\tau_3 y^\ga (\p_{1\ga} +\p_{2\ga}) +
z^\ga (\tau_2 \p_{2\ga}-\tau_1 \p_{1\ga}) +i(\tau_3+\tau_4) \p_{1\ga}\p_2^\ga
+ i\bar \p_{1\dga}\bar \p_2^\dga\,,
\ee
\be
\p_{i\ga} : =\f{\p}{\p y_i^\ga}\q \bar \p_{i\dga} := \f{\p}{\p \bar y_i^\dga}\,.
\ee
It is important that since $\tau_4\geq 0$, and hence $\tau_3+\tau_4 \geq \tau_3$,
the expansion coefficients in powers
of $\p_{1\ga}\p_2^\ga$ are larger than those of $y^\ga \p_{i\ga}$, which as will be
explained later, indicates nonlocality.

Let us now use Schouten identity
\be
z_\ga y^\ga \p_{1\gb} \p_2^\gb   -z^\ga\p_{1\ga} y^\gb \p_{2\gb}  +z^\ga \p_{2\ga}
y^\gb \p_{1\gb} =0\,
\ee
expressing the fact that antisymmetrization over any three two-component indices is zero to transform $B_{2\eta}$ to a different form. To this end, we observe that
 \be
  i\p_{1\ga} \p_2^\ga \exp (X)
= \f{\p}{\p \tau_4} \exp (X)\,,
 \ee
\be
  z^\ga \p_{1\ga} \exp (X)
= -\f{\p}{\p \tau_1} \exp (X)\,,
 \ee
\be
  z^\ga \p_{2\ga} \exp (X)
= \f{\p}{\p \tau_2} \exp (X)\,,
 \ee
\be
 \Big ( z_\ga y^\ga +i y^\ga (\p_{1\ga} + \p_{2\ga}) \Big )\exp (X)
= i\Big (\f{\p}{\p \tau_3} - \f{\p}{\p \tau_4}\Big ) \exp (X)\,.
 \ee
Using these relations and integrating by parts we obtain
\bee
 && 0=-\int d_+^4\tau \delta (1-\sum^4_{i=1}\tau_i)
\Big ( z_\ga y^\ga \p_{1\gb} \p_2^\gb   -z^\ga\p_{1\ga} y^\gb \p_{2\gb}  +z^\ga \p_{2\ga}
y^\gb \p_{1\gb} \Big ) \exp (X)\\
&&
\ls=\int d_+^4\tau \Big ((iy_\ga z^\ga \delta(\tau_4) +y^\gb (\p_{2\gb}\delta(\tau_1) +\p_{1\gb}\delta(\tau_2))
 \delta (1-\sum^4_{i=1}\tau_i)
+(\delta(\tau_3)-\delta(\tau_4)) \delta'(1-\sum^4_{i=1}\tau_i) \Big ) \exp (X)\,.\nn
\eee
Comparison of this with (\ref{B2}) allows us to represent $B_{2\eta}$ in the form
\bee
\label{B22}
 B_{2\eta} &&= -\f{1}{2}\eta \int d_+^4\tau \Big (
 iy_\ga z^\ga \delta(\tau_4)
 \delta (1-\sum^4_{i=1}\tau_i)
+(\delta(\tau_3)-\delta(\tau_4)) \delta'(1-\sum^4_{i=1}\tau_i)\Big )\nn\\
&&\exp (X)
 C(Y_1;K)C(Y_2;K) \Big \vert_{Y_{1,2}=0} *k*\kappa \,.
\eee
The terms with $\delta(\tau_3)$ and $\delta(\tau_4)$ have different meaning.
The part with $\delta (\tau_3)$ is $z$-independent. Indeed, denoting it
$\Delta C_{2\eta}$ we obtain using (\ref{uf}) and (\ref{[uf]})
\bee
\label{C2}
 \Delta C_{2\eta} &&\ls= -\f{1}{2}\eta \int d_+^4\tau \delta(\tau_3)
 \delta'(1-\sum^4_{i=1}\tau_i)\nn\\
&&\ls\exp (y^\ga (\tau_1 \p_{1\ga}-\tau_2 \p_{2\ga}) +i\tau_4 \p_{1\ga}\p_2^\ga
+ i\bar \p_{1\dga}\bar \p_2^\dga)
 C(Y_1;K)C(Y_2;K) \Big \vert_{Y_{1,2}=0} *k \,.
\eee

Since $\Delta C_{2\eta}$ is in the $\dr_Z$ cohomology, according to (\ref{eq:gen_sol})
we can define a new resolution operator
\be
\label{shift}
\dr^*_{loc} B := \dr^*_Z B - \Delta C_{2\eta}\,
\ee
 such that
 \be
 B^{loc}_{2\eta}= -\f{i}{2} \dr^*_{loc} ([S_1\,, C]_*)
 \ee
 has the form
 \be
 B^{loc}_{2\eta}=
 \f{1}{2}\eta \int d_+^3\tau \Big (
 \delta'(1-\sum^3_{i=1}\tau_i)-iy_\ga z^\ga
 \delta (1-\sum^3_{i=1}\tau_i) \Big )\exp (X^{loc})
 C(Y_1;K)C(Y_2;K) \Big \vert_{Y_{1,2}=0} *k*\kappa \,,
 \ee
 \be
X^{loc}=  i(1-\tau_3) z_\ga y^\ga +\tau_3 y^\ga (\p_{1\ga} +\p_{2\ga}) +
z^\ga (\tau_2 \p_{2\ga}-\tau_1 \p_{1\ga}) +i\tau_3 \p_{1\ga}\p_2^\ga
+ i\bar \p_{1\dga}\bar \p_2^\dga\,.
\ee
Equivalently, $ B^{loc}_{2\eta}$ can be represented in the integral form
\bee
\label{bloc}
B^{loc}_{2\eta} &&\ls = \f{1}{2}\eta \int d_+^3\tau
\big(\delta' (1-\sum^3_{i=1}\tau_i) -iy_\ga z^\ga
 \delta (1-\sum^3_{i=1}\tau_i)\big)
\int d^4 U d^4 V\exp i(U_A V^A +(1-\tau_3) z_\ga y^\ga)
\nn\\
&&
 C(\tau_3 y -\tau_1 z +\tau_3 u,\bar y+\bar u;K)
C(\tau_3 y+\tau_2 z +v,\bar y +\bar v;K)
*k*\kappa \,.
\eee
Formula (\ref{shift}) just describes the nonlinear shift  found in \cite{Vasiliev:2016xui}
to reduce the nonlocal bilinear corrections to the local form directly in the sector
of $x,y$-variables.

We observe that $B^{loc}_{2\eta}$ (\ref{bloc}) has the remarkable property that
the coefficient in front of the term  responsible for the index contraction between the first and second
factors of $C$ equals to those in front of the $y$ variable in the arguments of $C$,
which is analogous to the star product of $Z$-independent functions.

The results of this paper prove that this is  the only option consistent with locality.
Indeed, we have shown that the dependence on $y$  in the arguments of $C$
 in (\ref{bloc}) contains the same dependence on the homotopy parameter as
 $u$, that determines the contractions of spinorial indices. This has to be
 compared with the original (non-local) source (\ref{bn}) where the coefficients
 in front of $u$ (or $v$) determining contractions are larger that those in
 front of $y$. Clearly, since $\tau_4\geq 0$
 the solution (\ref{bloc}) is strictly minimally nonlocal, having the same type of
 nonlocality as the original star product in $y$ variables.

 The following comments are now in order.

 An important feature of $B^{loc}_{2\eta}$ (\ref{bloc}) is that it contains the  rightmost
 star-product factor $k*\kappa$.  Because, by (\ref{uf}), (\ref{[uf]}), star product with $\kappa$ exchanges $y$ and $z$ variables, Eq.~(\ref{bloc}) can be equivalently rewritten in the form
\bee
\label{bloc1}
B^{loc}_{2\eta} &&\ls = \f{1}{2}\eta \int d_+^3\tau
\big(\delta' (1-\sum^3_{i=1}\tau_i) +iy_\ga z^\ga
 \delta (1-\sum^3_{i=1}\tau_i)\big)
\int d^4 U d^4 V\exp i(U_A V^A +\tau_3 z_\ga y^\ga)
\nn\\
&&
 C( \tau_1 y - \tau_3 z +\tau_3 u,\bar y+\bar u;K)
C(-\tau_3 z-\tau_2 y +v,\bar y +\bar v;K)
*k \,.
\eee
 In this form the coefficient in front of $u$ that governs contractions is
 dominated by that
 in front of $z_\ga y^\ga$ in the exponential, as well as the coefficients in
 front of $z$ in the arguments of $C$. We observe that
 the  relevant terms  disappear
 at $Z=0$, not allowing to distinguish between proper and improper nonlinear corrections
 in attempt to analyze the issue of locality at $Z=0$ as  in
 \cite{Boulanger:2015ova,Skvortsov:2015lja}. This feature is in agreement
 with the well-known fact that the overall coefficients in front of different currents
 are not determined by the lower-order consistency and hence one can seemingly freely
 go from one set of coefficients to another by a nonlocal field redefinition of the
 original variables which are those associated with the $Y$ dependence at $Z=0$.
Such arguments led some of the authors of  \cite{Boulanger:2015ova,Skvortsov:2015lja}
 to claims that it is impossible to compute current vertices from the
  HS  equations of \cite{Vasiliev:1992av}. (See e.g.
 \cite{Sleight:2017pcz} for such interpretation of conclusions
 of \cite{Skvortsov:2015lja}).  In fact, the meaning of the results of
\cite{Skvortsov:2015lja} is that the authors were
using a bad luck {\it ad hoc} assumption that the problem can be
analyzed with the help of the conventional resolution operator $\dr^*_Z$ (\ref{eq:dz*}).
To make a proper choice,
 higher-order effects have to be taken into account.

 Indeed, our results imply
 that the setup of \cite{Vasiliev:2016xui} not only leads to the local result
 in the first nontrivial approximation but, most significantly,  it will lead
 to the minimally nonlocal setup in the higher orders. To see this it is important to
 have expressions for the bilinear corrections that account for  the  full  dependence
 on both $Y$ and $Z$. The computation of higher-order corrections will involve
 star products of the expressions like $B^{loc}_{2\eta}$ (\ref{bloc}) or
 $B_{2\eta}$ (\ref{bn}) with $B_{2\eta}$ being more nonlocal than $B^{loc}_{2\eta}$.
 Also let us note that  the factor of $k*\kappa$ is central and involutive.
 Hence it will cancel in particular the similar factor on the \rhs of (\ref{SS})
 in the next order with the effect that no exchange of $y$ and $z$ variables will
 occur in  (some of) the higher orders, \ie the minimal order of nonlocality will
 be visible directly in the physical $Y$-space in the higher-order terms. From this
 perspective our approach is somewhat similar to the derivation of cubic HS vertices
 by Metsaev in  \cite{Metsaev:1991mt} where the proper form of cubic HS vertices that
 precisely corresponds \cite{Gelfond:2017wrh,Misuna:2017bjb} to that resulting from
 the application of the resolution $\dr^*_{loc}$ to equations of \cite{Vasiliev:1992av}
 was deduced from the higher-order analysis.

One of the main results of this paper is the identification of the proper
resolution operator $d_{loc}^*$ (\ref{shift}) that directly leads to the formulation of the
HS theory in the local (or minimally nonlocal in the higher orders) setup
for the full system of nonlinear HS equations. So far, $d_{loc}^*$ was found
only in the lowest order of the 0-form sector. The goal is to find its full fledged
extension to all orders and all types of differential forms. This is the ongoing
project \cite{hom}.

Completion of the latter project will
also imply the completion of the program of \cite{Vasiliev:2015wma} of establishing a
proper class of star-product functions associated with the minimally nonlocal
setup in HS equations. Indeed, in \cite{Vasiliev:2015wma} the proper dependence on
$Z$ and $Y$ variables was established for the expressions like
\be
\label{ffreg}
f(Z;Y)=\int_0^1 d\tau \phi( \tau Z;(1-\tau)Y;\tau)
\exp{i\tau Z_A Y^A}\,
\ee
 with
$\phi( W; U;\tau)$  regular in $W$ and $U$ and integrable in $\tau$.
Being accompanied by the factor of $\tau$ and $1-\tau$, the
dependence on $Z$ and $Y$ on the {\it r.h.s.} of (\ref{ffreg})  trivializes
at $\tau \to 0$ and $\tau\to 1$, respectively. In
\cite{Vasiliev:2015wma} the space of functions (\ref{ffreg}) called
 $\bV_{0,0}$ was extended  to
the spaces $\bV_{k,l}$ of such star-product elements
(\ref{ffreg}) that $\phi(W;U;\tau)$ scales as $\tau^k$ at $\tau\to 0$ and  $(1-\tau)^l$
at $\tau\to 1 $. More precisely, we allow (poly)logarithmic dependence
on $\tau$ and $1-\tau$ at $\tau\to 0$ and $\tau\to 1$, respectively, with the
convention that it does not affect the indices $k$ and $l$.
In  \cite{Vasiliev:2015wma}   $\bV_{k,l}$ with both  positive and negative $k$ and/or $l$ were considered.

The problem not considered in \cite{Vasiliev:2015wma} was which restrictions on
the inner structure of  $\phi( W;U;\tau)$ have to be imposed to respect locality
or minimal nonlocality. In particular, a question  to be addressed is what
is the proper dependence on $\tau_i$ in the expressions like

\be
\label{cl}
  \int d_+^3\tau \rho(\tau)
\int d^2 u d^2 v\exp i(u_\ga v^\ga +t z_\ga y^\ga)
 C(\tau_3 y -\tau_1 z +\tau_5 u)
C(\tau_4 y+\tau_2 z +v)
*k*\kappa \,.
\ee
The answer combining the results of \cite{Vasiliev:2015wma} with those of this
paper is
\be
\label{t}
\tau_3\leq 1-t\q \tau_4\leq 1-t\q\tau_1\leq t \q\tau_2\leq t\q \tau_5\leq 1-t\,,
\ee
where the  new restriction is the last one dominating the dependence
on $\tau_5$ by that on the parameter $t$ in front of the factor of $z_\ga y^\ga$
in the exponential.

Let us stress that not only solutions have to be of the form (\ref{cl}), (\ref{t})
but also gauge transformation and  field redefinitions (as explained in
\cite{Vasiliev:2015wma}, the latter should even obey
stronger  conditions due to further restrictions on the measure $\rho(\tau)$).
In any case, the $Z$-independent field redefinition (\ref{C2}) has $t=1$
and hence does not belong to the proper class. Moreover, that $\tau_5\leq 0$
for $Z$-independent functions obeying (\ref{t})
implies that they
should be distributions  supported
at $\tau_5=0$ hence being represented by a finite number of
delta-function derivatives
$\delta^n(\tau_5)$. In turn this means that the allowed class of
$Z$-independent field redefinitions is genuinely local.
 Direct analysis of the next section also demonstrates that
$\dr^*_Z$ and $\dr_{loc}^*$ are not  related by a local
field redefinition.

\section{Cohomology shift and Green function}
\label{cur0}
In this section we show that the cohomology shift (\ref{C2}) relating $\dr_Z^*$ and
$\dr_{loc}^*$ is nonlocal. To this end we first recall in Section \ref{y} the structure
of the nonlinear corrections to dynamical equations in the sector of $Y$ variables
(\ie $\dr_Z$ cohomology) resulting from $\dr_Z^*$,  computing the zero-form Green
function in Section \ref{gf0}.

\subsection{$\dr_Z^*$-induced nonlocal deformation in the zero-form sector  }
\label{y}

{The lowest-order deformation
of  free equations (\ref{CON1}), (\ref{CON2}) has the form
  \be\label{defgo}
  \dr\go+ \go*\go+\Ll(w,\go,C)+ \G_{cur}(w ,\PPP)=0\,,
\ee
\be
\label{defc}  \dr C+ \go*C -C*\go+\Hh_{cur}(w,\PPP)=0\,,
\ee
where $\Ll(w,\go,C)$ is at most linear in both $\go$ and $C$ while
the two-form $\G_{cur}(w ,\PPP)$ and one-form $\Hh_{cur}(w,\PPP)$ are some functionals of the background fields $w$
and the current $\PPP$
\be
\label{JCC}
\PPP(Y_1,Y_2;K|x) := C(Y_1;K|x)  C(Y_2;K|x)\,.
\ee
As a consequence of equations (\ref{CON2}) on $C$, so defined current obeys the {\it current equation}
\be\label{tw2}
 \left( D_{L}   -{i} h^{\ga\pb}\Big (y^1{}_\ga \bar{y}^1{}_\pb- y^2{}_\ga \bar{y}^2{}_\pb
 -\p_1{}_\ga\bp_1{}_\pb+\p_2{}_\ga\bp_2{}_\pb\Big )\right)
 \PPP (y^1{},y^2{};\bar y^1{},\bar y^2{};K|x)=0
\ee
 at the convention that
 derivatives $\p_{1\ga}$($\bar \p_{1\dga}$) and $\p_{2\ga}$($\bar \p_{2\dga}$)
over the first and second  undotted(dotted)
spinorial arguments of $\PPP$ are defined to anticommute with
$k(\bar k)$.
The  star product in (\ref{defgo}), (\ref{defc})
results from the restriction of (\ref{star2}) to $Z$-independent functions.

Gauge invariant current interactions are associated with
$\G_{cur}(w ,\PPP)$ and $\Hh_{cur}(w,\PPP)$ linear in $\PPP$.
Other $\go$-dependent terms  bilinear in fluctuations, describe
gauge non-invariant interactions which do not contribute if the
spins $s_1$ and $s_2$ of two fields entering the bilinear terms on the
right-hand sides of (\ref{defgo}) and (\ref{defc}) and spin of the current
$s_\PPP$ identified with the spin of the field contributing to the linear
part of (\ref{defgo}) and (\ref{defc}) (\ie by its definition, the current of spin $s_j$
contributes to the nonlinear corrections to the equations on the spin-$s_\PPP$
field) obey the condition
\be
\label{sss}
s_\PPP\geq s_1+s_2\,.
\ee
(For the derivation of (\ref{sss}) directly
from Eq.~(\ref{defc}) see \cite{Gelfond:2015poa}).

The  final result of \cite{Vasiliev:2016xui} in the 0-form sector is
\bee
\label{F}
\Hh_{cur}(w,\PPP) &&\ls=-\f{i}{2}\int_0^1 d\tau
\Big (\eta
\int\f{ d\bar s d\bar t}{(2\pi)^{2}} \exp i [ \bar s_\dgb  \bar t^\dgb ]
 h(y, \tau\bar s +(1-\tau) \bar t)
 \PPP(\tau y , -(1-\tau) y  , \bar y +\bar s , \bar y +\bar t)\nn\\
&&
+\bar\eta \int \f{d s d t}{(2\pi)^{2}}
 \exp i [  s_\gb   t^\gb ] h(\tau  s -(1-\tau) t,\bar y)
\PPP(y+s,y+t,\tau \bar y,- (1-\tau) \bar y )\Big )
\,,
\eee
where
\be
\label{h}
h(u,\bar u) = h^{\ga\dga} u_\ga \bar u_\dga\,.
\ee
This corresponds to $\Hh_{cur}(w,\PPP)$ in
(\ref{defc}) with the minimal number of derivatives which is finite for any
 spins $s_1$, $s_2$  and $s_\PPP$.

The deformation $\Hh (w,\PPP)$ resulting from the nonlinear equations of
\cite{Vasiliev:1992av} by virtue of the resolution operator $\dr^*_Z$ is
\cite{Boulanger:2015ova,Vasiliev:2016xui}
\be
\label{hh}
\Hh (w,\PPP) = \Hh_\eta (w,\PPP)+\Hh_{\bar{\eta}} (w,\PPP)\,,
\ee
where
\bee
\label{CC}
\Hh_\eta (w,\PPP) =&&\ls-\f{i}{2}  \eta\int\f{ dS dT}{(2\pi)^{4}}
\exp i S_A T^A \int^1_0 d\tau \nn\\
&& [ h(s, \tau \bar y - (1-\tau) \bar t)
\PPP( \tau s,-(1-\tau)y +t; \bar y +\bar s,\bar y+\bar t;K)\nn\\
&&- h (t, \tau \bar y  - (1-\tau)\bar s) \PPP
((1-\tau)  y +  s, \tau  t, \bar y+ \bar s;\bar y+\bar t;K)]*k \,,
\eee
\bee
\label{barCC}
\Hh_{\bar\eta} (w,\PPP) =&&\ls-\f{i}{2} \bar \eta \int \f{dS dT}{(2\pi)^{4}}
\exp i S_A T^A \int^1_0 d\tau
\nn\\
&&\ls\ls[  h (\tau  y  - (1-\tau) t,\bar s) \PPP
(y+s,y+t; \tau \bar s, -(1-\tau)\bar y +\bar t;K)\nn\\
&&\ls\ls- h ( \tau  y  - (1-\tau) s,\bar t) \PPP
(y+s,y+t;(1-\tau) \bar y + \bar s, \tau \bar t);K]\bar *k\,.
\eee

The integration over $S$ and $T$ in (\ref{CC}), (\ref{barCC})  brings infinite tails
 of contracted indices, inducing by (\ref{CON2}) and (\ref{tw})
an infinite expansion in higher space-time derivatives of the constituent
fields. Hence,  $\Hh$ (\ref{CC}), (\ref{barCC}) differs from the conventional
current interactions  (\ref{F}) which, being free of the integration over $s_\ga$
and $t_\ga$, contains a finite number of derivatives for any  $s_1$, $s_2$
 and $s_J$.

To reproduce standard current interactions from those resulting from $\dr^*_Z$
one has to find a field redefinition
\be
\label{red}
C \to C'(Y;K|x) = C(Y;K|x)+ \Phi (Y;K|x)
\ee
with $\Phi$  linear in $\PPP$, bringing $\Hh$ (\ref{hh}) to the form (\ref{F}), \ie
\be
\label{hdh}
\Hh_{\eta} (w,\PPP) =  D_0 (\Phi_{\eta})
+ \Hh_\eta {}_{\,cur} (w,\PPP)\,.
\ee
The proper field redefinition found in \cite{Vasiliev:2016xui} is $\Phi =
\Delta C_{2\eta} + \Delta C_{2\bar \eta}$ with  $\Delta C_{2\eta}$ (\ref{C2})
  and $\Delta C_{2\bar \eta}$ being its  complex conjugate.
It should be stressed that
formula (\ref{hdh}) is valid for any spins $s_1$, $s_2$ and $\s_\PPP$ independently
on whether the condition (\ref{sss}) holds or not.

\subsection{Green function}
\label{gf0}
To discuss locality properties of the field redefinition (\ref{red})
it is useful to consider Green function that
removes the current interactions.
The goal is to find a solution to the equation
\be
DG_C (\PPP) =\Hh_\eta {}_{\,cur} (w,\PPP)\ + \Hh_{\bar \eta}{}_{\,cur} (w,\PPP)\,.
\ee
Since $\Hh_\eta {}_{\,cur} (w,\PPP)$ and $\Hh_{\bar \eta}{}_{\,cur} (w,\PPP)$
describe an arbitrary local current with certain coefficients the resulting solution $G_C(\PPP)$ describes
Green function  applied to  $\PPP$. Let us stress that this problem setting
is only applicable at the condition (\ref{sss}), when the HS connections $\go$
do not contribute. So, we will assume that  (\ref{sss}) is true.
Since the problem is linear,
the terms proportional to $\eta$ and $\bar \eta$ can be found separately. We consider
the part $G_{\eta}(\PPP)$ linear in $\eta$. The term $G_{\bar \eta}(\PPP)$
with $\bar \eta$ can be obtained by complex conjugation.

Let us look for $G_{\eta}(\PPP)$ in the most general Lorentz covariant form
\be
\label{Gppp}
G_{\eta}(\PPP) =  \phi (N_i, \bar N_i, M,\bar M) \PPP(Y;\bar Y;k,\bar k)
\Big \vert_{Y_1=Y_2=0}\,,
\ee
where
\be
\label{N}
N_i = y^\ga\p_{i \ga } \q \overline N_i = \bar y^\dga \bar\p_{i\dga }\,,
\ee
\be
\label{M}
M= \epsilon^{\ga\gb}\p_{1 \ga} \p_{2 \gb}\q
\overline M= \epsilon^{\dga\dgb}\bar \p_{1 \dga} \bar \p_{2 \dgb}\,,
\ee
\be
\p_{i \ga}:= \frac{\p}{\p y_i^\ga}\q \bar \p_{i\dga} := \frac{\p}{\p \bar y_i^\dga}\,.
\ee
We use  convention that both $y^\ga$ and $y_i^\ga$ along with the respective
derivatives anticommute with the Klein operator $k$ inside $\PPP$, while
$\bar y^\dga$ and $\bar y_i^\dga$  anticommute with $\bar k$.

Using (\ref{CON2}) and (\ref{tw2}) it is not difficult to obtain
\bee
\label{DG}
DG_\eta(\PPP) &&\ls =i\lambda h^{\ga\dgb} \Big \{y_\ga \bar y_\dgb
\Big (-1 +\frac{\p^2}{\p N_1\p \overline N_1} +\frac{\p^2}{\p N_2\p \overline N_2}\Big ) -
y_\ga {\bar \p_{1\dgb}} \frac{\p^2}{\p N_2\p \overline M}+
y_\ga \bar \p_{2\dgb} \frac{\p^2}{\p N_1\p \overline M}\nn\\
&&
-\p_{1 \ga} \bar y_\dgb \frac{\p^2}{\p M\p \overline N_2}
+\p_{2\ga} \bar y_\dgb \frac{\p^2}{\p M\p \overline N_1}+
\p_{i\ga} \bar \p_{j\dgb} \frac{\p^2}{\p N^i \p \overline N^j}\nn\\
&&
- \Big(\p_{1\ga}\p_{1 \dgb}  +\p_{2\ga}\bar \p_{2\dgb}
\Big)
\Big (1 -\frac{\p^2}{\p M\p \overline M}\Big)\Big \}\phi (N_i, \bar N_i, M,\overline M)
  \PPP(Y_1;Y_2;K)
 \Big \vert_{Y_1=Y_2=0} \,.
\eee

To reproduce $\Hh_{\eta \,cur} (w,\PPP)$ (\ref{hh}) we have to demand that all terms
on the r.h.s. of (\ref{DG}) should be zero except for those containing
$y_\ga \bar \p_{i\dgb}$. However, demanding this, one should take into account
that antisymmetrization over any three two-component indices gives zero. This yields the
relations
\be
\label{Fierz}
M y_\ga + N_2 \p_{1\ga} - N_1 \p_{2\ga} =0\q
\overline M \bar y_\dga + \overline N_2 \bar\p_{1\dga} - \overline N_1 \bar\p_{2\dga} =0\
\ee
allowing to add the following expression to $DG_\eta$ (\ref{DG})
\bee
\label{O}
O(J)&&\ls= i\lambda h^{\ga\dgb} [(M y_\ga +N_2 \p_{1\ga} -N_1 \p_{2\ga})
(\bar a \bar y_\dgb+\bar b_1\bar\p_{1\dgb} +\bar b _2\bar \p_{2\dgb})\nn\\
&& + (ay_\ga +b_1 \p_{1\ga} +b_2 \p_{2\ga}) (\overline M\bar y_\dgb +
\overline N_2\bar \p_{1\dgb} - \overline N_1 \bar \p_{2\dgb})]\PPP(Y_1;Y_2;K)
 \Big \vert_{Y_1=Y_2=0}  \,,
\eee
where $a$, $b_i$, $\bar a$ and $\bar b_i$ are arbitrary
functions of $N_i,M,\overline N_i$ and $\overline M$. As a result,
\bee
\label{DG}
DG_\eta(\PPP) &&\ls +O =i\lambda h^{\ga\dgb} \Big \{y_\ga \bar y_\dgb
\Big (-\phi +\frac{\p^2\phi}{\p N_1\p \overline N_1} +
\frac{\p^2\phi}{\p N_2\p \overline N_2}+ M\bar a +a \overline M\Big )\nn\\
&& + y_\ga {\bar \p_{1\dgb}}\Big(
- \frac{\p^2 \phi}{\p N_2\p \overline M}+ M \bar  b_1 +\overline N_2 a\Big )+
y_\ga \bar \p_{2\dgb}\Big (\frac{\p^2\phi}{\p N_1\p \overline M} +M\bar b_2 -
\overline N_1 a \Big )
\nn\\
&&
+\p_{1 \ga} \bar y_\dgb \Big (-\frac{\p^2\phi}{\p M\p \overline N_2} +
\overline M b_1 +N_2 \bar a\Big )
+\p_{2\ga} \bar y_\dgb \Big (\frac{\p^2\phi}{\p M\p \overline N_1} +\overline M  b_2
-N_2\bar a\Big )\nn\\
&&
+
\p_{1\ga} \bar \p_{1\dgb}\Big ( \frac{\p^2 \phi}{\p N_1 \p \overline N_1}
+\frac{\p^2\phi}{\p M\p \overline M}-\phi +N_2 \bar b_1 +\overline N_2 b_1\Big )
\nn\\
&&
+
\p_{2\ga} \bar \p_{2\dgb}\Big ( \frac{\p^2 \phi}{\p N_2 \p \overline N_2}
+\frac{\p^2\phi}{\p M\p \overline M}-\phi -N_1 \bar b_2 -\overline N_1 b_2\Big )
\nn\\
&&
+
\p_{1\ga} \bar \p_{2\dgb}\Big (\frac{\p^2 \phi}{\p N_1 \p \overline N_2} +N_2 \bar b_2
-\overline N_1 b_1 \Big )
+
\p_{2\ga} \bar \p_{1\dgb}\Big (\frac{\p^2 \phi}{\p N_2 \p \overline N_1} +
\overline N_2  b_2
- N_1 \bar b_1 \Big )
\Big \} \PPP(Y_1;Y_2;K)
 \Big \vert_{Y_1=Y_2=0} \,.
\eee

Now we can solve the equation
\be
\label{DGO}
DG_C (\PPP) +O(\PPP) =\Hh_\eta {}_{\,cur} (w,\PPP)\,
\ee
 ignoring relations (\ref{Fierz}). However, this equation admits a solution
 only at the condition (\ref{sss}) since otherwise it is simply inconsistent
 as long as the contribution of the HS connections $\omega$ is not taken into account.

 To project the currents to appropriate helicities $h_J$, $h_1$ and $h_2$, where
 \be
 \label{hhh}
2 h_\PPP = y^\ga \f{\p}{\p y^\ga} - \bar y^\dga \f{\p}{\p \bar y^\dga}\q
2h_i =   y_i^\ga \f{\p}{\p y_i^\ga} - \bar y_i^\dga \f{\p}{\p \bar y_i^\dga}
 \ee
 the function $\phi$ should be chosen  appropriately. Firstly, we observe that
 the operators
 \be
 x_i = N_i \overline N_i\q z:= M\overline M
 \ee
do not affect the helicities since they commute with the operators (\ref{hhh}).

Ansatz that is consistent with condition (\ref{sss}), allowing to solve (\ref{DGO}), is
\be
\label{phiphi}
G_{C\,s_1,s_2,s_\PPP} = N_1^{n_1} N_2^{n_2} \overline M^t \varphi(x_i,z)\,,
\ee
where, assuming for definiteness that $s_1\geq s_2$
(the opposite case can be considered analogously)
\be
\label{nnt}
n_1=s_j+s_1+s_2\q n_2=s_j-s_1-s_2\q t=s_j-s_1+s_2\,.
\ee
This implies, in particular,
\be
\label{sts}
2s_1\geq t\geq 2s_2\,
\ee
and
\be
h_1=s_1\q h_2 = -s_2\,.
\ee
The complex conjugated Ansatz solves for $\Hh_{\bar \eta}{}_{\,cur} (w,\PPP)$.
In this paper we only consider the case of opposite helicities $h_1 h_2\leq0$
which is sufficient for our purposes. It would be interesting to extend the obtained
results to helicities of equal signs as well. In particular, this is useful for the
analysis of holography along the lines of \cite{Didenko:2017lsn}.

It is also convenient to use  the following Ansatz
\be
a(N,M)=N_1^{n_1} N_2^{n_2} \overline M^{t-1} \ga(x,z)
\ee
\be
\bar a(N,M)=N_1^{n_1} N_2^{n_2} \overline M^{t+1} \bar\ga(x,z)
\ee
\be
b_1(N,M)=N_1^{n_1} N_2^{n_2+1} \overline M^{t} \gb_1(x,z)
\ee
\be
b_2(N,M)=N_1^{n_1+1} N_2^{n_2} \overline M^{t} \gb_2(x,z)
\ee
\be
\bar b_1(N,M)=N_1^{n_1} N_2^{n_2+1} \overline M^{t} \bar \gb_1(x,z)
\ee
\be
\bar b_2(N,M)=N_1^{n_1-1} N_2^{n_2} \overline M^{t} \bar \gb_2(x,z)
\ee
Plugging this into (\ref{DGO}) gives the following system of differential
equations in the variables $x_i$ and $z$
\be
\label{1}
-\varphi +( n_1 +1 + x_1 \p_{x_1})\p_{x_1} \varphi + (n_2+1+x_2 \p_{x_2})\p_{x_2}\varphi
+\ga -z \bar \ga =0\,,
\ee
\be
\label{2}
(n_2+x_2 \p_{x_2})(t+z\p_z) \varphi -z\bar\gb_1 -x_2 \ga = \psi_1(x)\,,
\ee
\be
\label{3}
(n_1+x_1 \p_{x_1})(t+z\p_z) \varphi -z\bar\gb_2 -x_1 \ga = \psi_2(x)\,,
\ee
\be
\label{4}
\p_{x_2}\p_z \varphi +\gb_1 +\bar  \ga = 0\,,
\ee
\be
\label{5}
\p_{x_1}\p_z \varphi -\gb_2 +\bar  \ga = 0\,,
\ee
\be
\label{6}
(n_1 +1 +x_1 \p_{x_1}) \p_{x_1} \varphi -\varphi +(t+1+z\p_z)\p_z +\bar \gb_1 -x_2 \gb_1=0
\,,
\ee
\be
\label{7}
(n_2 +1 +x_2 \p_{x_2}) \p_{x_2} \varphi -\varphi +(t+1+z\p_z)\p_z +\bar \gb_2 +x_1 \gb_2=0
\,,
\ee
\be
\label{8}
(n_1 +x_1 \p_{x_1})\p_{x_2} -\bar \gb_2 +x_1 \gb_1 =0
\,,
\ee
\be
\label{9}
(n_2 +x_2 \p_{x_2})\p_{x_1} -\bar \gb_1 -x_2 \gb_2 =0
\,,
\ee
where
\be
\p_{x_i}:=\frac{\p}{\p x_i}\q \p_z:=\frac{\p}{\p z}\,.
\ee
Here $\psi_1 (x)$ are functions of $x_i$ which should
reproduce $\Hh_\eta {}_{\,cur} (w,\PPP)$ (\ref{hh}).
$\psi_1 (x)$ are demanded to be independent of $z$ since,
containing no integration over $s$ and $t$
$\Hh_\eta {}_{\,cur} (w,\PPP)$ (\ref{hh}) contains no
contractions of undotted indices. In Appendix we shall see that it is enough
to demand this to derive the proper form of $\psi_1 (x)$ that corresponds to
$\Hh_\eta {}_{\,cur} (w,\PPP)$ by solving the system (\ref{1})-(\ref{9}).
The final result is
\bee
\label{varphixxzser}
G_{\eta}(\PPP) = &&\ls
   \eta \sum_{r_{1,2}, \bar r_{1,2}, p, \bar p=0}^\infty
 \theta (r_1-\bar r_1) \theta (r_2-\bar r_2) \theta (\bar p -p)
\theta( r_1 -\bar r_1 -\bar p +p) \theta( \bar p - p  - r_2+\bar r_2)\nn
\\&&\times\f{i^{\bar p-p+1}}
{\,\bar r_1 !\, \bar r_2!\,\bar p!\,
(r_1 +r_2 +p+1)!}N_1^{r_1}N_2^{r_2}\overline N_1^{\bar r_1}
\overline N_2^{\bar r_2} M^p \overline M^{\bar p} \PPP(Y_1,Y_2;k,\bar k)
\Big \vert_{Y_1=Y_2=0}
\,.
\eee

On the other hand, using  the generalized beta-function formula
\cite{Gelfond:2017wrh}
 \bee\label{inttau1-tau01pp}
 \int\! d\gt^m \gd^{(k)} \!\left(\!1\!-\!\sum\limits_{i=1}^m\gt_i\!\right)
  \prod\limits_{i=1}^m\theta(\gt_i)\gt_i^{n_i}  =
 \f{ \prod\limits_{i=1}^m  {n_i}! }{ \Big(\!\sum\limits_{i=1}^m n_i +m \!-\!1\!-\!k\!\Big)!}
 \q\forall n_i, k\ge0 \,\eee
the field redefinition (\ref{shift}) can be rewritten in the form
\be
\label{varloc}
\Phi_{\eta}(\PPP) =   \eta \sum_{r_{1,2}, \bar r_{1,2}, p, \bar p=0}^\infty i^{\bar p-p+1}
 \f{1}
{\,\bar r_1 !\, \bar r_2!\,\bar p!\,
(r_1 +r_2 +p+1)!}N_1^{r_1}N_2^{r_2}\overline N_1^{\bar r_1}
\overline N_2^{\bar r_2} M^p \overline M^{\bar p} \PPP(Y_1,Y_2;k,\bar k)
\Big \vert_{Y_1=Y_2=0}
\,.
\ee
We observe that the Green function (\ref{varphixxzser}) differs from the field redefinition
(\ref{shift}) only by the factors of $\theta$ restricting the field redefinition to the
 region (\ref{sss}), (\ref{nnt}), (\ref{sts}).

The fact that in the allowed region of helicities the Green function
(\ref{varphixxzser}) and  field redefinition (\ref{varloc}) coincide is
not accidental. It is a  consequence of the property that, as is easy to see,
 the contribution (\ref{CC}) resulting from application of the
 resolution $\dr^*_Z$ to the nonlinear equations is zero in this sector.
Hence, in this sector, the field redefinition (\ref{varloc}) must have the form of the
 Green
function applied to the resulting local current. This means that the field redefinition
(\ref{shift}) is essentially nonlocal, \ie in agreement with the conclusions of
Section \ref{loc} the resolution operators $\dr^*_Z$ and
$\dr^*_{loc}$ are locally nonequivalent.

\section{Discussion}
\label{disc}
We have identified the proper resolution operator in the space of spinorial $Z^A$-variables,
 that leads to local
 first nonlinear correction to  HS equations. It is shown to
correspond to certain class of functions of the type identified in \cite{Vasiliev:2015wma}
extended to the terms accounting contractions between indices of the field product factors.
As shown in \cite{Didenko:2017lsn}, the current interactions of \cite{Vasiliev:2015wma}
resulting from this resolution operator properly reproduce the
anticipated HS holographic results.

It should be stressed that the distinguished role of the resolution $\dr_Z^*$
becomes manifest only if the dependence on $Z^A$ affecting
 the higher-order nonlinear corrections to HS equations  is taken into account.
It is however  less obvious at $Z^A=0$ in which case relevant terms in the deformation
trivialize.

Naive interpretation of our results might be that
the field redefinition (\ref{varloc}) relating the simplest resolution operator
$\dr_Z^*$ in the HS theory to  $\dr_{loc}^*$
 is not allowed, being nonlocal. The proper interpretation however
 is just opposite: to reach a minimally nonlocal setup in HS theory (which is fully
  local to the order in question), one
 has to  use the resolution  $\dr_{loc}^*$ with no reference to $\dr_Z^*$
 at all. It is $\dr_{loc}^*$ that fulfils locality compatible  boundary conditions
 in the process of solving HS equations with respect to $Z^A$-variables. Hence,
 the proper interpretation is that $\dr_Z^*$ is related to the local resolution
 $\dr_{loc}^*$ by a nonlocal field redefinition making the $\dr_Z^*$ setup  improper from
 the  locality perspective.

 This raises the question of the proper definition of the resolution operator
  $\dr_{loc}^*$ at the higher orders and its extension to
 the sector of one-forms. These issues will be considered in \cite{hom}. The
 analysis of higher orders is also interesting in the context of
 conclusions of the recent paper  \cite{Sleight:2017pcz} claiming that
 the level of nonlocality in HS gauge theories is somewhat extreme.
 This remains to be analyzed carefully from several perspectives,
however.

One is that the space-time derivatives in $AdS$ do not commute,
having the commutator of order one in dimensionless units
$[\lambda^{-1} D\,, \lambda^{-1} D]\sim 1$. This raises the
question in which  ordering prescription the properties of the
functions of the covariant D'Alambertian $f(\Box)$ are analyzed.
As is well known, going from one ordering  to another
may significantly affect analytic properties of the function in question.
For instance, being exponentials in the HS star product (\ref{star2}),
inner Klein operators have a form of distributions in the Moyal-Weyl star product
\cite{Didenko:2009td} which property in fact highlights the distinguished role
of the HS star product (\ref{star2}) in the HS gauge theory.

Another is that in presence of an infinite tower of HS states
even local field redefinitions at the level of quadratic corrections may induce
nonlocal contributions at higher orders. This phenomenon  has to be properly
 taken into account in the
locality analysis of the cubic corrections to the HS field equations.

Also it should be stressed that the relation between the form of
local corrections in terms of spinor $Y^A$-variables and that in terms
of space-time derivatives via (\ref{tw})   acquires nonlinear corrections.
As a result,  analysis of the problem in terms of spinors
may have much simpler form, simultaneously providing a distinguished
ordering prescription mentioned above.

Finally,  as a byproduct of our consideration we have found explicit expression for
the zero-form Green function in the case  of the
constituent fields of  opposite  helicity signs. It would be interesting to extend these results to
the helicities of the same sign as well as to the sector of  one-forms.

\section*{Acknowledgements}
I am grateful to Slava Didenko, Olga Gelfond and Massimo Taronna for useful discussions and comments.
This research was supported by the Russian Science Foundation Grant No 14-42-00047.
I would like to thank the Galileo Galilei Institute for Theoretical Physics (GGI)
for the hospitality and INFN for partial support during the completion of this work.
 This work also was partially supported by a grant from the Simons Foundation

\newcounter{appendix}
\setcounter{appendix}{1}
\renewcommand{\theequation}{\Alph{appendix}.\arabic{equation}}
\addtocounter{section}{1} \setcounter{equation}{0}
 \renewcommand{\thesection}{\Alph{appendix}.}
 \addtocounter{section}{1}

\section*{Appendix}
The system (\ref{1})-(\ref{9}) contains nine equations on
seven arbitrary functions $\ga$, $\gb_i$, $\bar \ga$, $\bar \gb_i$
and $\varphi$ of $x_i$ and $z$.  In fact, only five independent
combinations of $\ga$, $\gb_i$, $\bar \ga$, $\bar \gb_i$ enter the system
since $O$ (\ref{O}) is invariant under the following transformations
\be a\to a +M u\q b_1\to b_1 + N_2 u \q b_2\to b_2 -N_1 u
\ee
\be
\bar a \to \bar a - \overline M u\q \bar b_1 \to \bar b_1 -\overline N_2 u\q
\bar b_2 \to \bar b_2 + \overline N_1 u\,
\ee
with arbitrary $u$.

Solving equations (\ref{1}), (\ref{4}), (\ref{5}), (\ref{6}) and (\ref{7})
one obtains
\be
\ga -z \bar \ga= \varphi -( n_1 +1 + x_1 \p_{x_1})\p_{x_1} \varphi - (n_2+1+x_2 \p_{x_2})\p_{x_2}\varphi
 =0\,,
\ee
\be
\gb_1 = -\bar \ga -\p_{x_2}\p_z \varphi\q \gb_2 = \bar \ga +\p_{x_1}\p_z \varphi\,,
\ee
\be
\bar \gb_1 =\varphi  -(n_1 +1 +x_1 \p_{x_1}) \p_{x_1}\varphi  -(t+1+z\p_z)\p_z \varphi -
x_2\bar\ga -x_2\p_{x_2} \p_z \varphi\,,
\ee
\be
\bar \gb_2 =\varphi  -(n_2 +1 +x_2 \p_{x_2}) \p_{x_2}\varphi  -(t+1+z\p_z)\p_z \varphi -
x_1\bar\ga -x_1\p_{x_1} \p_z \varphi\,.
\ee

Plugging this into the remaining equations gives four equations on $\varphi$ while,
as anticipated, the dependence on $\bar \ga$ drops out. Namely, Eqs.~(\ref{2}), (\ref{3})
yield
\be
\label{1'}
(n_2+x_2\p_{x_2} +z\p_z)(t+x_2\p_{x_2} +z\p_z)\varphi -(z+x_2)(\varphi - (n_1 +x_1\p_{x_1} +1)\p_{x_1}
\varphi)
=\psi_1(x)\,,
\ee
\be
\label{2'}
(n_1+x_1\p_{x_1} +z\p_z)(t+x_1\p_{x_1} +z\p_z)\varphi -(z+x_1)(\varphi - (n_2 +x_2\p_{x_2} +1)\p_{x_2}
\varphi)
=\psi_2(x)\,,
\ee
while Eqs.~(\ref{8}) and (\ref{9}) yield
\be
\label{3'}
\varphi -(n_1+n_2 +x_1\p_{x_1} +x_2\p_{x_2} +1) \p_{x_2}\varphi -
(t+z\p_z+1)\p_z \varphi +x_1(\p_{x_2}- \p_{x_1})\p_z \varphi =0\,,
\ee
\be
\label{4'}
\varphi -(n_1+n_2 +x_1\p_{x_1} +x_2\p_{x_2} +1) \p_{x_1}\varphi -
(t+z\p_z+1)\p_z \varphi +x_2(\p_{x_1}-\p_{x_2}) \p_z \varphi =0\,.
\ee
Now we are in a position to solve these equations for $\varphi$ which is not hard
because the system (\ref{1'})-(\ref{4'}) is largely overdetermined.

The difference between (\ref{3'}) and (\ref{4'}) gives
\be
\big (n_1+n_2 +x_1\p_{x_1} +x_2\p_{x_2}-(x_1+x_2) +1\big) (\p_{x_1}-\p_{x_2})
\varphi(x_1,x_2,z) =0\,.
\ee
For $\varphi(x_1,x_2,z)$ expandable in power series of $x_i$ and $z$ this implies that
\be
\label{x}
\varphi(x_1,x_2,z) = \tilde  \varphi(x,z)\q x:=x_1+x_2\,.
\ee
Plugging this back into (\ref{3'}) yields
\be
\label{5'}
\tilde \varphi -(n_1+n_2 +x\p_{x}  +1) \p_{x}\tilde \varphi -
(t+z\p_z+1)\p_z \tilde \varphi  =0\,.
\ee
This equation gives
\be
\tilde \varphi = \f{1}{(n_1+n_2 +x\p_x)!(t+z\p_z)!}\,e^{x} \chi(w)\q w=x-z\,,
\ee
where $\chi(w)$ is an arbitrary function of a single variable. Plugging this
into (\ref{1'}), (\ref{2'}) one finds that the necessary condition, that the
left-hand sides of these equations are $z$-independent, is
\be
\p_w [(n_1 +n_2 +w\p_w +w +1)\chi(w)]=0\,.
\ee
This equation is solved by
\be
\label{chi}
\chi(w) =  \chi_0\f{(w\p_w)!}{(n_1+n_2 +w\p_w +1)!} e^{-w}\,,
\ee
where $\chi_0$ is a constant. This yields
\be
(n_1 +n_2 +w\p_w +w +1)\chi(w) = \f{1}{(n_1+n_2)!}\chi_0\,.
\ee
Finally, plugging (\ref{chi}) into equations (\ref{1'}) and (\ref{2'})
after some transformations one finds that they are indeed solved provided that
\be
\label{psi1}
\psi_1 = \chi_0 \f{(n_2+x_2\p_{x_2})}{(t-1)! (n_1+n_2+x\p_x +1)! (n_1+n_2)!} \,e^x\,,
\ee
\be
\label{psi2}
\psi_2 = \chi_0 \f{(n_2+x_1\p_{x_1})}{(t-1)! (n_1+n_2+x\p_x +1)! (n_1+n_2)!}\, e^x\,.
\ee
These  reproduce (\ref{hh}) provided that
\be
\chi_0 =i^{t+1} \eta (n_1+n_2)!\,.
\ee

This determines the Green's function in the form (\ref{Gppp}), (\ref{phiphi}) with
\be
\label{varphixxz}
\varphi(x_1,x_2,z) = i^{t+1}\eta \f{(n_1+n_2)!}{(n_1+n_2 +x\p_{x})!}
\exp{x} \f{(x\p_x+z\p_z)!}{(n_1+n_2 +x\p_x+z\p_z+1)!(t+z\p_z)!} \exp{(z-x)}\,.
\ee
Using that
\be
\int_0^1 d\tau \tau^n (1-\tau)^m =\frac{n!m!}{(n+m+1)!}
\ee
this expression can be further evaluated as
\bee
\label{varphixxz}
\varphi(x_1,x_2,z) &&\ls= i^{t+1}\eta \int_0^1 d\tau \tau^{n_1+n_2}
\exp{x} \f{1}{(n_1+n_2 +x\p_{x})!(t+z\p_z)!} \exp{(1-\tau)(z-x)}\nn\\
&&\ls=  i^{t+1}\eta \int_0^1 d\tau \tau^{n_1+n_2}
 \f{1}{(n_1+n_2 +x\p_{x})!(t+z\p_z)!} \exp{(\tau x +(1-\tau)z)} \,.
\eee

Note that if  conditions (\ref{nnt}) are not true,
equations (\ref{1})-(\ref{9}) admit no polynomial solutions at all because some of
involved factorials will diverge, \ie the obtained formulae
hold only in the chosen area of helicities.
Plugging this expression into (\ref{phiphi}) we obtain
(\ref{varphixxzser}). This does not mean however that the Green function cannot
be constructed in the case of helicities $h_1$ and $h_2$ of the same sign in which case
being formally nonpolynomial function of its arguments, the proper solution is
anticipated to be regular in the allowed region of spins (\ref{sss}).

\end{document}